# Scaling nanowire-supported GaN quantum dots to the sub-10-nm-limit, yielding complete suppression of the giant built-in potential

Swagata Bhunia[1,2], Ritam Sarkar[2], Dhiman Nag[2], Dipankar Jana[3], Suddhasatta Mahapatra [1, *], Apurba Laha [2, *]

1. Department of Physics, Indian Institute of Technology Bombay, Mumbai-400076, India
2. Department of Electrical Engineering, Indian Institute of Technology Bombay, Mumbai-400076, India
3. Department of Condensed Matter Physics & Materials Science, Tata Institute of Fundamental Research, Mumbai-400005, India

**Abstract:** The nanowire-supported quantum dot (NWQD) of GaN is an unconventional nanostructure, which is extremely promising for realization of UV photonics in general, and room-temperature single photon generation, in particular. While GaN-NWQDs have several promising attributes, the crucial challenge in exploiting their full potential, is to reduce the lateral dimensions of the QDs, to the order of the exciton Bohr-radius in GaN. Also critical is to suppress the built-in electric field due to spontaneous and piezoelectric polarization, which adversely affects the radiative recombination lifetime. We report here the innovation of a simple yet powerful single-step epitaxial growth technique, to achieve both of these targets. By combining controlled and on-demand thermal decomposition of GaN nanowires, with our previously-developed strategy of inhibiting the same via AlN-capping, we demonstrate that the NWQD-diameter can indeed be reduced to the truly strong-quantum-confinement limit. In these ultra-scaled GaN QDs, we show that the built-in electric fields are almost completely suppressed. The NWQD fabrication-strategy developed in this work may pave the way for fabrication of highly efficient classical and quantum UV-emitters based on GaN.

Introduction:

In recent years, nanowire-supported-quantum-dots of Gallium Nitride (GaN) have emerged as a unique building block for both classical and quantum light emission in the UV region of the electromagnetic spectrum [1-5]. GaN-based quantum dots (QDs), in general, have long been in focus of UV-photonics research, owing to the high chemical and thermal stability [6], large exciton binding energy [7], and high-dielectric breakdown voltage [8] of GaN, and its large band-offsets with other III-Nitride semiconductors [9]. The large band offsets, and a large hole-effective-mass, yields a higher degree of quantum confinement in GaN QDs, in comparison to that provided by their III-Arsenide or II-VI counterparts [5]. When compared to GaN quantum wells, the stronger confinement in GaN-QDs not only suppresses defect-state-mediated carrier-leakage [10], but also reduces the effect of the large polarization field of the material, thus yielding faster radiative recombination [11-13].

While GaN QDs show such promising attributes, their fabrication continues to be challenging. For applications requiring control over size, position, and areal density, QDs self-assembled in the Stranski-Krastanov (SK) growth mode of heteroepitaxy are not very suitable [14, 15]. Moreover, SK QDs of GaN are prone to formation of defect states, which leads to unwanted residual emission and spectral diffusion [16, 17]. Also, the 2D-wetting layer beneath the ensemble of the QDs, which is characteristic of the SK growth mechanism, provides a direct pathway for carrier-escape, at higher temperatures [18]. Finally, a strong built-in polarization field (of the order of MV/cm [19]), undermines the benefits of the strong quantum-confinement effect, in (0001)-oriented GaN QDs, typically obtained in the SK growth mode.

Forming QDs at the apices of nanowires (NWs), within a single epitaxial growth step, is an attractive alternative to the SK method, for obtaining high quality III-Nitride QDs [5, 20, 21]. Unlike the planer growth front available for self-assembly, the

NW provides a dislocation-free crystalline surface for subsequent QD-growth, and also offers a large surface-to-volume ratio [22, 23]. Furthermore, the approach enables deterministic positioning and control of the QD areal density, when suitably pre-patterned substrates are used [5, 21]. Another advantage of the approach is that the entire nanowire-quantum-dot (NWQD) structure can be transferred to a foreign substrate, thus allowing easier integration into optoelectronic and photonic circuits [24]. All these favorable properties are much-coveted for improving the device characteristics of GaN-based light-emitting sources. For example, very high internal quantum efficiency (IQE) and light extraction efficiency have been obtained for GaN-NWQD-based UV-LEDs (IQE > 90% has been reported at the single-QDs level) [25-29]. Secondly, the threshold current density of GaN-NWQD-based lasers has been observed to be significantly low [30]. Besides LEDs and lasers, high-temperature single-photon emission has also been successfully demonstrated with such QDs, which are characterized by a shorter exciton radiative lifetime and large energy separation between quantized states [31].

Although significant progress has been made towards realization of GaN-NWQD-based LEDs, lasers, and single-photon sources (SPS), the application-potential of this unique building-block is still not fully explored. This is because reducing the lateral dimension of the GaN NWQD, down to the excitonic Bohr-radius of GaN, has so-far proven to be non-trivial [21, 31]. When fabricated on the c-plane (0001) of NWs, either the QD-diameter remains identical to that of NW, or the QD-growth extends onto the sidewall of the NW [21, 31]. Moreover, similar to the situation of self-assembled SK QDs, a strong internal electric field persists also in (0001)-oriented GaN-NWQDs [32], which adversely affects the switching time and threshold current density of LEDs and laser diodes (LDs) [33]. Ge-doping of the NWQDs [34] has been observed to partially mitigate this issue, while the concept of Internal-Field-Guarded-Active-Region Design (IFGARD) has been developed recently, to obtain quasi-electric-field-free regions for NWQD growth [35]. However, the former approach degrades the crystal quality of the NWQDs, while with IFGARD, a certain fraction of the emitted photons is reabsorbed.

In this work, we report the innovation of a simple technique to obtain NWQDs with lateral dimensions comparable to the excitonic Bohr radius of GaN, by on-demand activation and suppression of thermal decomposition of GaN NWs, during molecular beam epitaxy (MBE) growth. Unlike in previous approaches for fabrication of NWQDs, sidewall-growth of the QDs is completely inhibited by this approach, thus allowing the vertical and the lateral dimensions of the QDs to be precisely controlled. Excitonic photoluminescence emission from these truly strongly-confined QDs reveals that the large built-in electric field, due to spontaneous and piezoelectric polarization in GaN QDs, is more-effectively suppressed by shrinking of the QD-diameter, rather than by reduction of the QD-height. We corroborate this finding with our single-band effective mass calculations, and demonstrate, for the first time, that the built-in polarization is almost completely suppressed, in the NWQDs fabricated by our approach.

Experimental methods:

The distinct strategy developed for realization of the GaN NWQDs is schematically shown in Figure 1. Thermal decomposition of GaN NWs at high temperatures (either during growth or during post-growth thermal annealing) is now well-established [36, 37]. In a previous report, we have shown that this decomposition can be completely suppressed by cresting the as-grown GaN NWs with a thin AlN cap layer [38]. Combining this capability with controlled decomposition of GaN NWs is the crucial enabling step of the NWQD formation technique, demonstrated here. Accordingly, GaN NWs of length ~ 550 nm and diameter ~ 58 nm were first grown, and crested with a ~ 3 nm AlN cap (fig. 1 (a)). Subsequently, GaN deposition was continued, which formed a ~ 150 nm GaN-extension of the AlN-crested GaN NWs, with the same diameter (fig. 1 (b)). The composite NW was then annealed at 950°C for 1 hour. This resulted in selective thermal-decomposition of only the 150-nm GaN extension, while the AlN-cap kept the bottom 550 nm GaN NW completely protected. The process yielded the "candle"-like structures shown in fig. 1 (c), consisting of a GaN "quantum wire", supported on the thicker base provided by the AlN-crested GaN NW. The QD was then positioned atop the GaN quantum wire (QW) by growing a tiny AlN/GaN/AlN heterostructure (fig. 1 (d)). For the chosen annealing duration of 1 hour, the QWs narrowed down to 16±2 nm, thus allowing QDs, with radius comparable to the exciton Bohr radius of GaN, to be achieved within the subsequently-deposited AlN/GaN/AlN heterostructure.

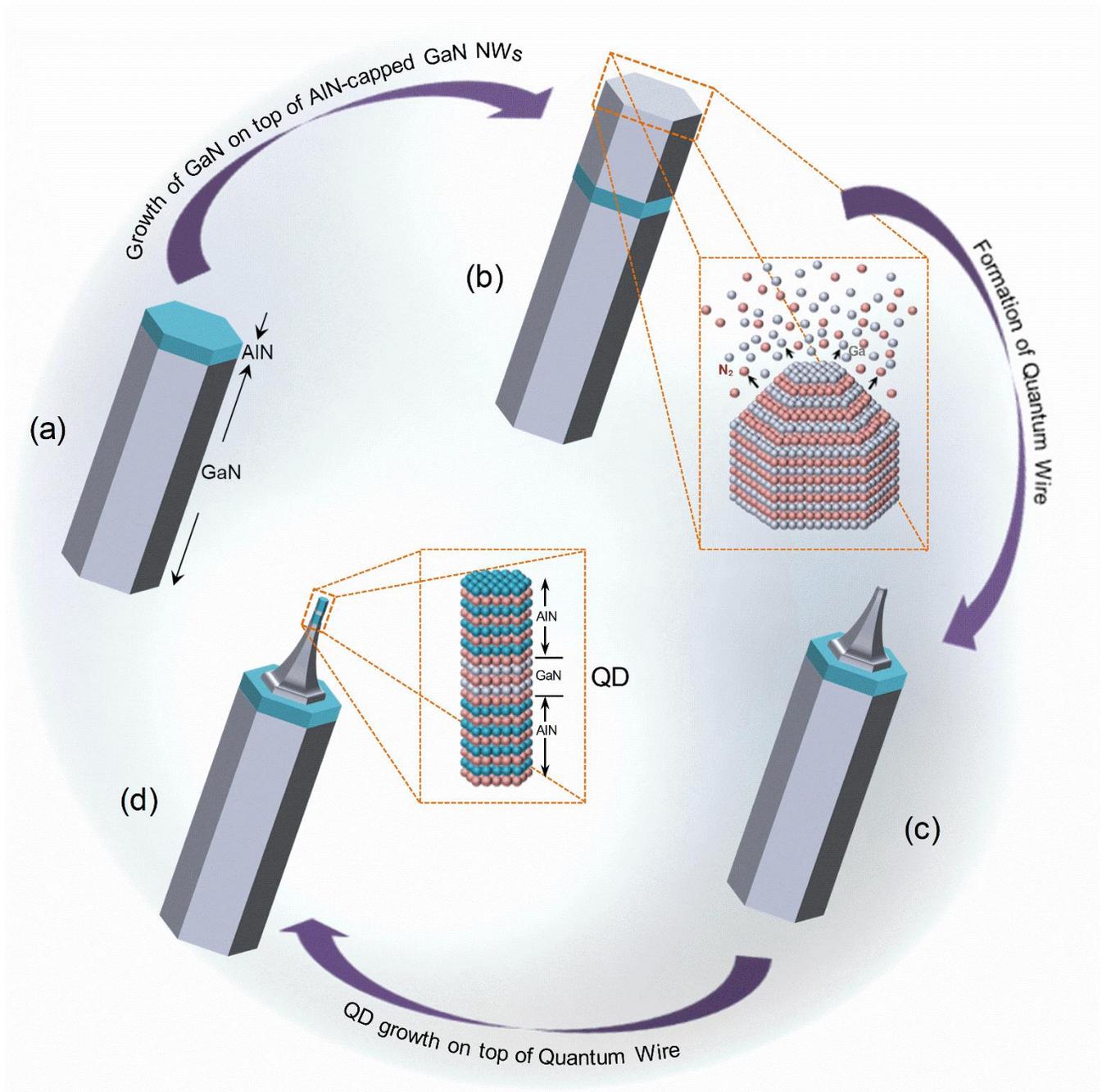

**Figure 1:** (a)-(d) Schematic representation of the key steps in the growth of GaN NWQDs. The GaN NW is first capped with AlN (a). Subsequently, GaN growth is continued to obtain the GaN-extension (b). The GaN QW is obtained on top of the AlN-capped GaN NW, after annealing at 950 °C (c). GaN thermal decomposition, selectively from the GaN-extension, leading to the formation of the QW, is shown in the inset of (b). The GaN QD is grown on top of the GaN QW, by growing a tiny AlN/GaN/AlN hetero-structure (d). A close-up schematic of the GaN QD, sandwiched between two AlN barriers, is shown in the inset of (d).

All samples were grown in a RIBER C21 molecular beam epitaxy (MBE) chamber, equipped with an ADD ON nitrogen plasma source and effusion cells for Al, Ga, In, Si and Mg. The details of growth of the AlN-crested GaN NWs have been reported in Ref. [38]. Growth of the first GaN NW, the AlN cap, and the subsequent GaN-extension, was carried out at 880 °C, with a $N_2$ flow rate and plasma forward power of 3.5 sccm and 475 W, respectively. The AlN/GaN/AlN heterostructure, defining the NWQD, was also deposited at the same temperature (after UHV-annealing at 950 °C for 1 hour). Here, we investigate three different NWQD samples, where the GaN deposition for the QD formation was carried out for 60 s, 40 s and 20 s, henceforth referred to as sample I, sample II, and sample III, respectively. Two other samples (IV and V) were prepared, where the AlN/GaN/AlN heterostructures were grown directly on top of the as-grown GaN NWs (without any annealing), to obtain dots of the same diameter as the NWs. For the fabrication of these large dots (henceforth referred to as quantum disks), the GaN deposition within the AlN/GaN/AlN heterostructures was carried out for 60 s and 30 s. These quantum-disk samples provide a premise for probing the effect of radial confinement in the true NWQD samples.

The growth of the samples was monitored real-time, using Reflection High Energy Electron Diffraction (RHEED). Field Emission Scanning Electron Microscopy (FESEM) and (High- Resolution) Transmission Electron Microscopy (HR-TEM) were used for characterizing the structural properties of all samples. Photoluminescence (PL) spectra were recorded from the samples at 10 K, using a CryLaS solid-state laser, with an excitation wavelength of 266 nm and a spot size of 300 µm.

Results and Discussions:

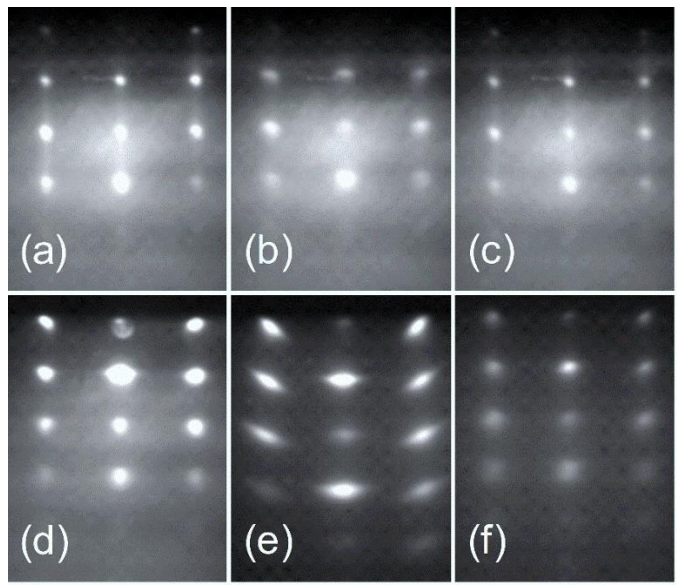

**Figure 2:** RHEED images recorded along the <11-20> azimuth of GaN, (a) after 150 min of GaN growth, (b) after 3 min of AlN-capping, (c) after 30 min of GaN growth on top of the AlN, (d) after annealing at 950°C for 30 min, and (e) after deposition of the AlN/GaN/AlN heterostructure. (f) RHEED image recorded along the <11-20> azimuth, after deposition of the AlN/GaN/AlN heterostructure, on as-grown GaN NWs (without annealing).

Formation of the QWs by the technique described earlier is already evidenced during growth, via RHEED. Figures 2(a) – 2(f) show the RHEED patterns along the (11-20) azimuth of GaN, captured at different stages of growth. The RHEED shows a spotty pattern after the growth of the GaN NWs (fig. 2(a)). The spotty pattern persists even after AlN capping of the GaN NWs (fig. 2(b)). However, the spots appear broader due to the proximity of the GaN and AlN reflections [38]. After subsequent growth of the GaN extension, on AlN-capped GaN NWs, the RHEED spots remain identical in size (fig. 2 (c)), to those of fig. 2(a). This implies that the diameter of the GaN extensions remain the same as that of the initially-grown GaN NWs. On the other hand, after annealing at 950°C, the RHEED spots become broader and brighter (fig 1(d)), indicating thinning of the extension regions due to thermal decomposition. Fig. 2 (e) shows the RHEED pattern recorded after growth of the AlN/GaN/AlN heterostructures, atop the GaN QWs. The distinctly spotty nature of the RHEED pattern implies that the entire structure remains fully crystalline, till the end of sample growth. However, the spots in fig. 2 (e) are observed to be slightly tilted. This tilting of the RHEED spots was not seen for growth of the AlN/GaN/AlN heterostructures directly on the GaN extensions, i.e. in the absence of the annealing step. Hence, it may be inferred that following the growth of the AlN/GaN/AlN heterostructures, the QWs are tilted away from the vertical axis [39].

A FESEM image, recorded from sample I, at a sample-tilt of 45°, is shown in fig. 3 (a). While the complete nanostructures are seen in this image, their apices are more discernible in the higher magnification image of fig. 3 (b). The images clearly depict the narrowing of the GaN-extensions (atop the AlN-capped GaN NWs) to form the QWs. This gives the distinct candle like shape to the structures, as mentioned earlier. The average length of the QWs is ~ 105 nm, while their diameter at the base is ~ 21 nm. The diameter at the tip of the QWs is too small to be reliably measured from these images. A significant number of the QWs are observed to be tilted by ~ 5°. This tilting of the QWs, evidenced earlier in RHEED images, may be attributed to the non-uniform AlN growth, during deposition of the QD-forming AlN/GaN/AlN heterostructure [39]. In the top-view FESEM image of the same sample (fig. 3 (c)), the spots with brighter contrast, at the center of the hexagonal cross-sections, are the QWs. Tilting of the QWs is also evident in this figure. Growth of such unique nanostructures, which has enabled the NWQD-diameter to be scaled down to the strong quantum-confinement limit, has not been reported before.

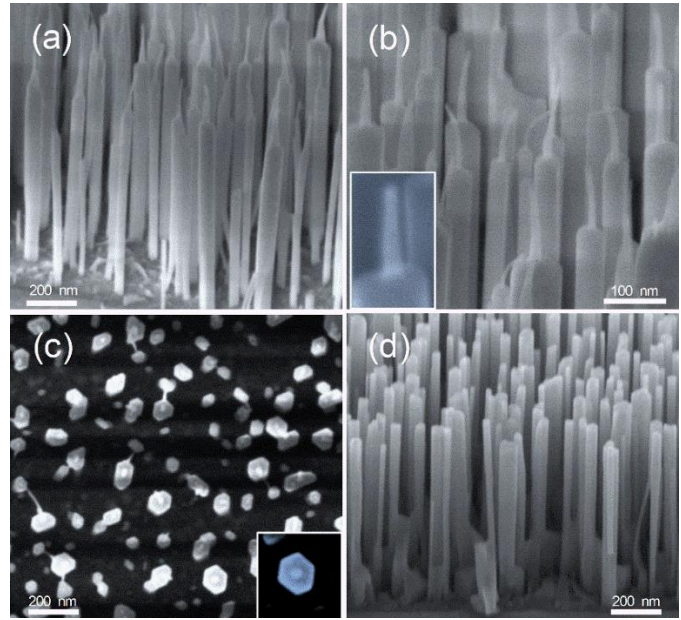

Figure 3: (a) FESEM image (recorded at 45°-tilt) of GaN QWs, supported by AlN-capped GaN NWs. (b) The apices of the same structures, showing both straight and tilted QWs more prominently. The inset shows a close-up of one such QW. (c) FESEM top-view of the structures, showing the presence of QWs (brighter spots) on top of the NWs. The inset shows the top view of a single QW on top of a single NW. A large difference between the diameters of the NW and the QW is clearly observed. (d) FESEM image (recorded at 45°-tilt) of uncapped GaN NWs.

On the other hand, the FESEM 45°-tilt image of a sample with only GaN NWs (without AlN-capping) shows a uniform cross-section, without any tapering or narrowing towards the apex (fig. 3 (d)). The length and diameter of the AlN-capped GaN NWs of sample I (measured by the ImageJ[40] software) are the same as those of the as-grown NWs of this sample (550 nm and 56 nm, respectively). This confirms that the AlN-cap protects the underlying GaN NWs from thermal decomposition, as demonstrated in our earlier work [38].

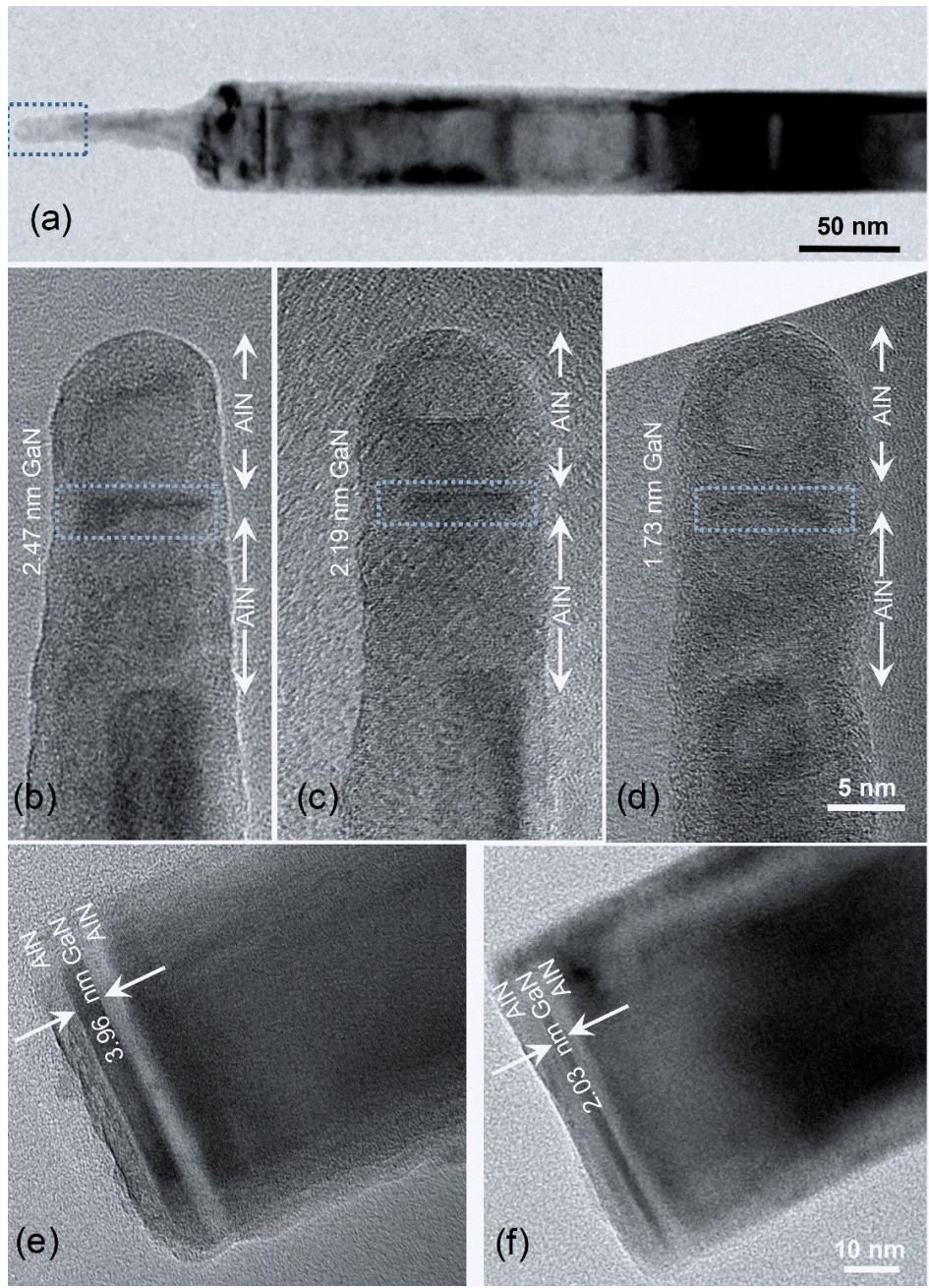

Figure 4: (a) TEM image of a complete NWQD structure. (b)-(d) HRTEM images of the GaN QWs (indicated by the box in (a)) within which the AlN/GaN/AlN QDs of different heights are embedded. (e), (f) HRTEM images of GaN quantum disks (embedded in AlN), of two different heights.

Figure 4 shows the TEM images of the NWQD samples I, II and III, together with samples consisting of quantum disks (samples IV and V). The whole candle-like structure is shown in fig. 4 (a). For all three samples, the height of the QWs is measured to be 104.75 nm, while their diameter at the tip is estimated to be 15 nm. Figs. 4 (b), 4 (c) and 4 (d) show the HRTEM images of the tip-regions of the QWs, in sample I, II and III, respectively. In these images, the GaN QDs, embedded within AlN barrier layers, can be clearly seen. Despite a 2.4 % lattice-mismatch between GaN and AlN, no threading dislocations can be traced within the QDs, or the barrier regions. This is because the free-surfaces along the QW-sidewalls provide a pathway for the lattice-mismatch-induced strain to be elastically relaxed. With GaN growth for 60 s (sample I), 40 s (sample II) and 20 s (sample III), within the AlN/GaN/AlN heterostructures, the obtained heights of the QDs are $h_{QD}^{(I)} = 2.47$ nm, $h_{QD}^{(II)} = 2.19$ nm and $h_{QD}^{(III)} = 1.73$ nm, respectively. On the other hand, the corresponding diameters of the QDs, $D_{QD}^{(I)} = 11.73$ nm, $D_{QD}^{(II)} = 11.21$ nm and $D_{QD}^{(III)} = 12.04$ nm, do not show any significant dependence on the duration of GaN growth. Since the excitonic Bohr radius of GaN is 3.0 nm, the lateral dimensions of the QDs obtained by this approach are in the strong quantum-confinement regime.

Another crucial aspect of the NWQDs formed by this technique is the absence of QD-growth on the sidewall surfaces, unlike as reported for previous approaches [31]. Unintentional side-wall growth makes reducing the QD-size to the true strong-confinement regime difficult. Moreover, the sidewall GaN deposit provides a path for carrier-escape, at high temperatures. By our formation technique, the sidewall-growth is completely suppressed since the growth temperature (880° C) is high, and the QW onto which the QD-heterostructure is deposited is already extremely thin. At 880° C, the decomposition rate of GaN from the sidewalls of these thin QWs is higher than the deposition rate.

Fig. 4 (e) and 4 (f) shows the HRTEM images of sample IV and sample V, respectively. Quantum disks of height 3.96 nm and 2.03 nm, and diameter of ~ 58.54 nm and 57.08 nm are observed for the two samples. As annealing was not performed in this case, the quantum disks are observed to be of similar diameter, as that of the initially-grown GaN NWs.

PL spectra recorded at 10 K, from the QD samples I, II and III, the quantum disk samples IV and V, and as-grown GaN NWs (without AlN capping), are shown in fig. 5. Excitonic emissions corresponding to sample I ($h_{QD}^{(I)} = 2.47$ nm, $D_{QD}^{(I)} = 11.73$ nm), sample II ($h_{QD}^{(II)} = 2.19$ nm, $D_{QD}^{(II)} = 11.21$ nm) and sample III ($h_{QD}^{(III)} = 1.73$ nm, $D_{QD}^{(III)} = 12.04$ nm) show peaks at $E_X^{(I)} = 3.639$ eV, $E_X^{(II)} = 3.739$ eV and $E_X^{(III)} = 3.845$ eV, respectively (figs. 5 (a) – 5 (c)). The peak positions were extracted by fitting the 10 K PL spectra with Gaussian distribution curves. The increasing blue-shift with reducing QD-height (and QD-diameter) is a clear signature of increasing quantum confinement. Taking sample I as an example, the blue-shift of the QD PL peak, with respect to that of the as-grown GaN NWs ($E_X^{(NW)} = 3.478$ eV, fig. 5 (d)), is 161 meV. However, the actual blue-shift due to strong lateral confinement in the QDs is estimated to be 659 meV. This is because the excitonic PL peak (at 10 K) of a GaN quantum well, of thickness (2.5 nm) comparable to the height of the QDs in sample I, appears at $E_X^{(QWell)} = 2.980$ eV [41]. Together with the PL from the QDs, the spectra in figs. 5 (a) –5 (c) also reveal emission peaks from the supporting GaN NWs (at 3.509 eV), and a broad-spectrum at 2.185 eV. The latter may be attributed to defect states at the GaN-AlN sidewall-interface of the QWs.

On the other hand, the quantum disks of sample IV ($h_{QDisk}^{(IV)} = 3.96$ nm, $D_{QDisk}^{(IV)} = 58.54$ nm) and sample V ($h_{QDisk}^{(V)} = 2.03$ nm, $D_{QDisk}^{(V)} = 57.08$ nm) show PL emission peaks at $E_X^{(IV)} = 2.950$ eV (fig 5(d)) and $E_X^{(V)} = 3.285$ eV (fig. 5 (e)). Since the diameters of the disks in both samples are almost same, the 335 meV blue-shift between samples IV and V directly reflects the increase in quantum confinement with decreasing height. More importantly, a comparison of the QD emission peak ($E_X^{(II)} = 3.739$ eV) of sample II ($h_{QD}^{(II)} = 2.19$ nm, $D_{QD}^{(II)} = 11.21$ nm) and the quantum disk emission peak ($E_X^{(V)} = 3.285$ eV) of sample V ($h_{QDisk}^{(V)} = 2.03$ nm, $D_{QDisk}^{(V)} = 57.08$ nm) reveals a blue-shift of 454 meV. This may be unambiguously ascribed to the ~ 5 times reduction in the diameter, since the heights of the QD and the quantum disk are approximately the same. However, as discussed next, the observed blue-shifts, with decreasing height and diameter of the emitters (both quantum disks and quantum dots), are a result of an intricate interplay between quantum confinement and built-in electric fields.

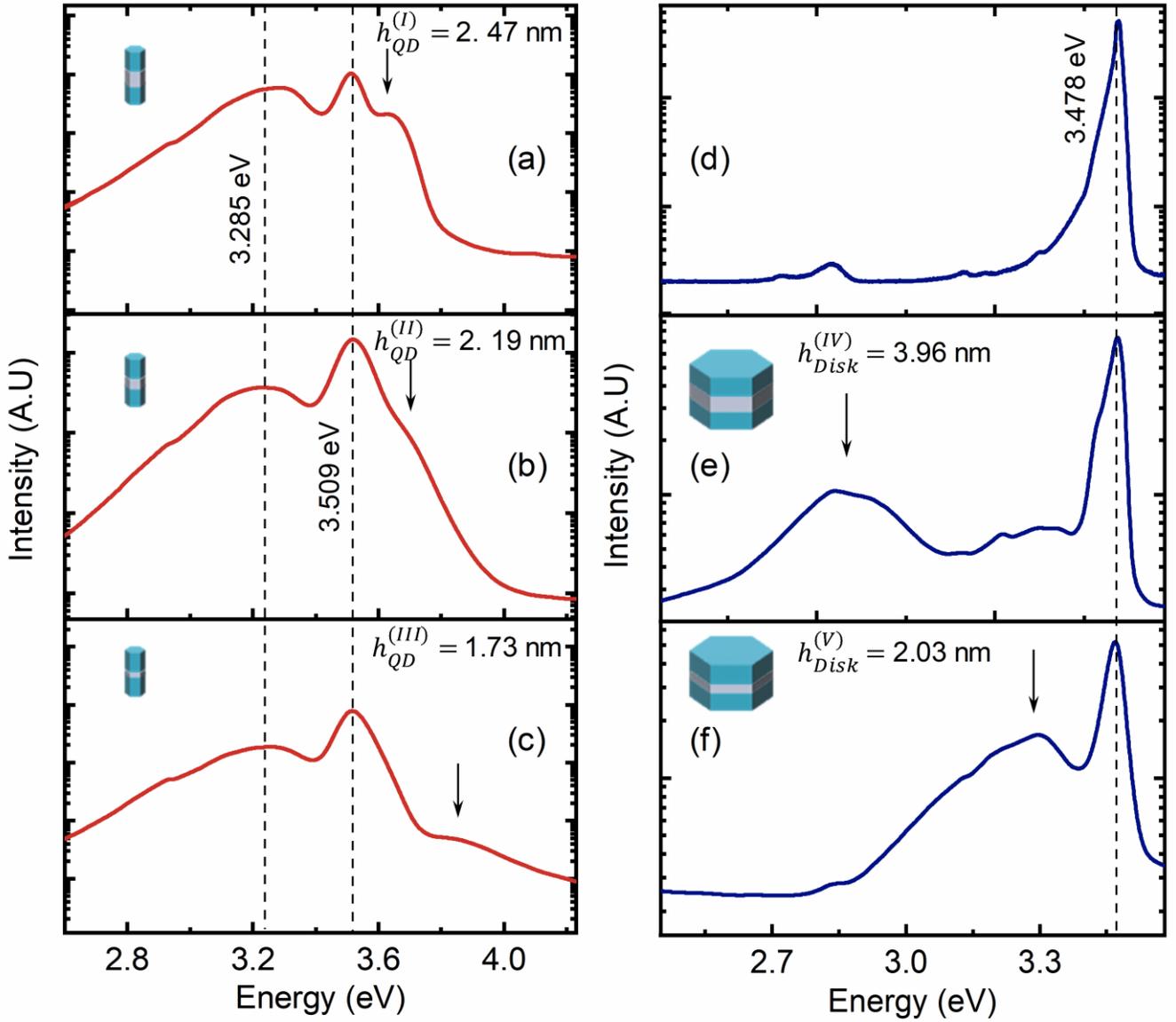

Figure 5: (a)-(c) PL spectra, recorded at 10 K, from NWQD-samples, with different QD-heights. The arrows indicate the emission peak from the QDs, which are blue-shifted with decreasing height of the QDs. (d) PL spectrum recorded at 10 K from the as-grown GaN NW sample. (e), (f) PL spectra, recorded at 10 K, from the quantum disk samples, with two different quantum-disk diameters. The arrows indicate the emission peaks from the quantum disks, which are also blue-shifted with decreasing height of the quantum disks.

In order to investigate the PL energy-shifts of the QDs, with variation of their height and diameter, we carried out effective mass calculations, with the envelope-function-approximation, using MATLAB. Instead of the 8-band k.p Hamiltonian, we used a one-band Hamiltonian, to reduce the computation time. This simplification is only expected to cause a shift of the ground state energies of the electron and the hole, by ~ 100 meV.

The one-band effective-mass Hamiltonian for electrons and holes are expressed as [42]

$H_{electron} = \Delta E_c(1 - \chi_{QD}) + \frac{\hbar^2}{2}\left[\frac{(k_x^2+k_y^2)}{m_c^{\parallel}} + \frac{k_z^2}{m_c^{\perp}}\right] + a_2(E_{xx} + E_{yy}) + a_1 E_{zz} - \Phi$

and

$$H_{hole} = -\Delta E_v(1 - \chi_{QD}) + (A_2 + A_4 - A_5)(k_x^2 + k_y^2) + (A_1 + A_3)k_z^2 + (D_2 + D_4 - D_5)(E_{xx} + E_{yy}) + (D_1 + D_3)E_{zz} - \Phi$$

Here, $\Delta E_c$ and $\Delta E_v$ are the barrier-heights seen by the electrons and holes, respectively, at the GaN/AlN heterojunction. $\chi_{QD}$, $E_{ij}$ ($i,j \in \{x,y,z\}$) and $\Phi$ are the QD-characteristic-function, the strain-tensor elements and the total built-in potential inside the QD, respectively. $A_i$, $D_i$ and $a_i$ are the Luttinger-like parameters, the valence-band deformation-potential and the conduction-band deformation-potential, respectively. $k_i$ ($i \in \{x,y,z\}$) are the wave vectors, and $m_c^\parallel$ and $m_c^\perp$ are the longitudinal and transverse electron effective-masses, respectively, in terms of the free electron mass.

The total built-in potential ($\Phi$) is the sum of the potential due to piezo-electric polarization ($\Phi_{PZ}$) and that due to spontaneous polarization ($\Phi_{SP}$). For obtaining analytical expressions for the piezoelectric and spontaneous polarization components, we have considered a truncated-pyramidal QD, with a hexagonal cross-section. The angle between the base and the side-facets of the pyramid is 45°. For such a QD, we calculated the strain-tensor elements in inverse Fourier space ($E_{ij}$ ($i,j \in \{x,y,z\}$) [43], and used them to obtain $\Phi_{PZ}$ [42]. Also $\Phi_{SP}$ was evaluated in inverse Fourier space, taking the difference of spontaneous polarization of GaN and AlN [42]. Further details of the calculation are given in supporting information.

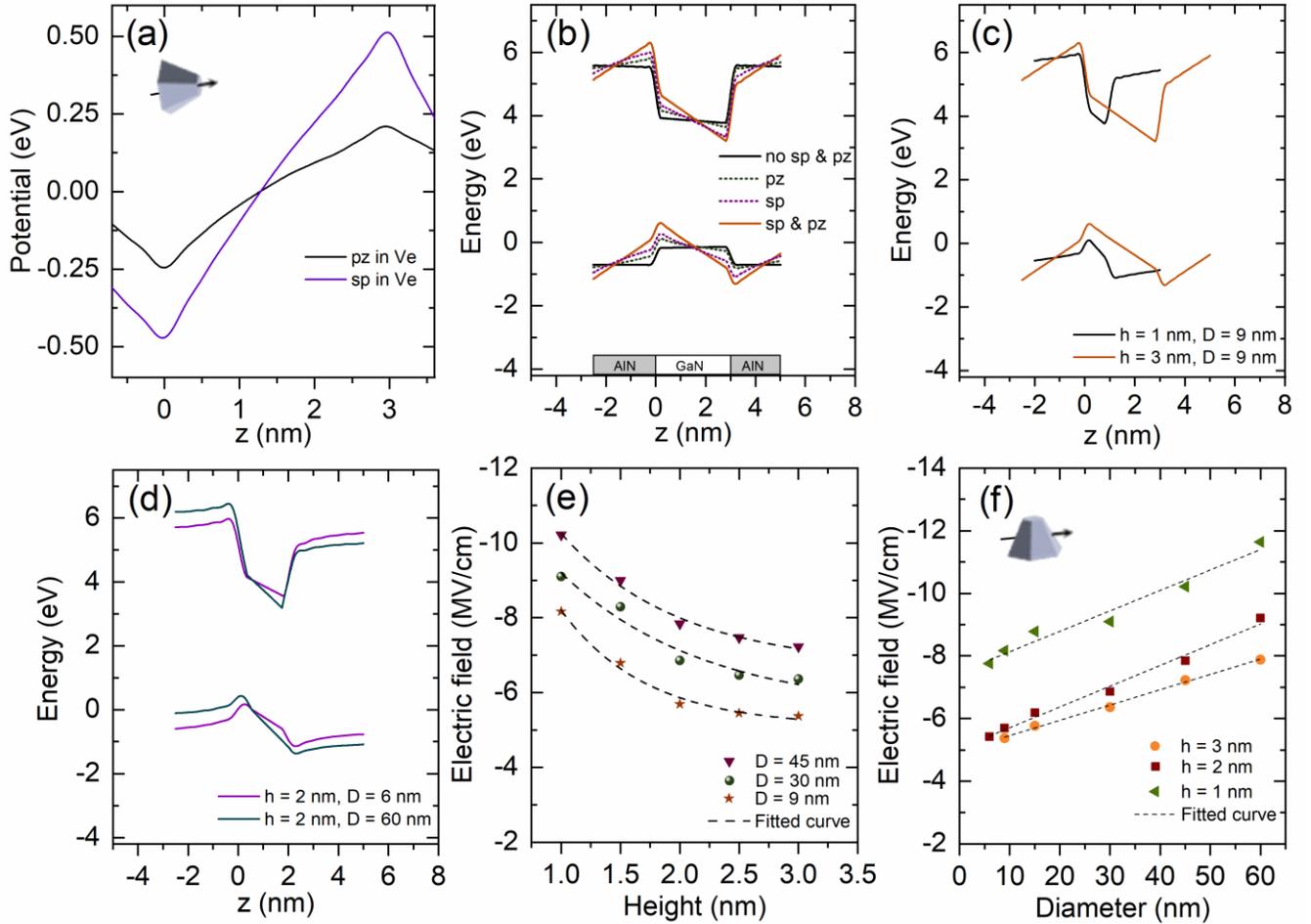

Figure 6 : (a) Built-in potential along the (0001) direction, due to piezo (pz) and spontaneous polarization (sp), plotted for a QD with a height of 3 nm, and a diameter of 9 nm. (b) The polarization-induced bending of the conduction and the valence bands, for the same QD. Polarization-induced band-bending, (c) for two different QD-heights, but the same QD-diameter, and (d) for two different QD-diameters, but the same QD-height. Plot of the variation of built-in electric field (e) with QD-height (for a definite QD-diameter) and (f) with QD-diameter (for a definite height).

Figure 6 (a) plots the variation of the built-in potential in the direction normal to the basal plane, both due to piezo-electric ($\Phi_{PZ}$) and spontaneous ($\Phi_{SP}$) polarization. Here, the QD-height is $h = 3$ nm, and the QD-diameter is $D = 9$ nm. Though $\Phi_{PZ}$ is smaller than $\Phi_{SP}$, its contribution to the total potential is non-negligible. The bending of the conduction and valence bands in the absence of $\Phi = \Phi_{PZ} + \Phi_{SP}$, and in the presence of only $\Phi_{PZ}$, only $\Phi_{SP}$ and $\Phi$ is shown in fig 6(b), for the same QD. As expected, the band-bending is strongest when the built-in potential due to both polarizations is considered. Fig. 6 (c) shows that the band-bending increases when $h$ is reduced (from 3 nm to 1 nm), keeping $D$ (= 9 nm) constant. On the other hand, Fig. 6 (d) shows that the band-bending decreases when $D$ is reduced (from 60 nm to 6 nm), keeping $h$ (= 2 nm) constant. This implies that the built-in potential, due to piezoelectric and spontaneous polarizations, decreases (increases) with reducing diameter (height) of the QDs. Fig. 6 (e) (Fig. 6 (f)), wherein the variation of the built-in electric field is plotted against $h$ ($D$), for different fixed values of $D$ ($h$), further consolidates this fact. As evident from these plots, the built-in electric field increases exponentially with decreasing height, while it reduces linearly with shrinking diameter of the QD.

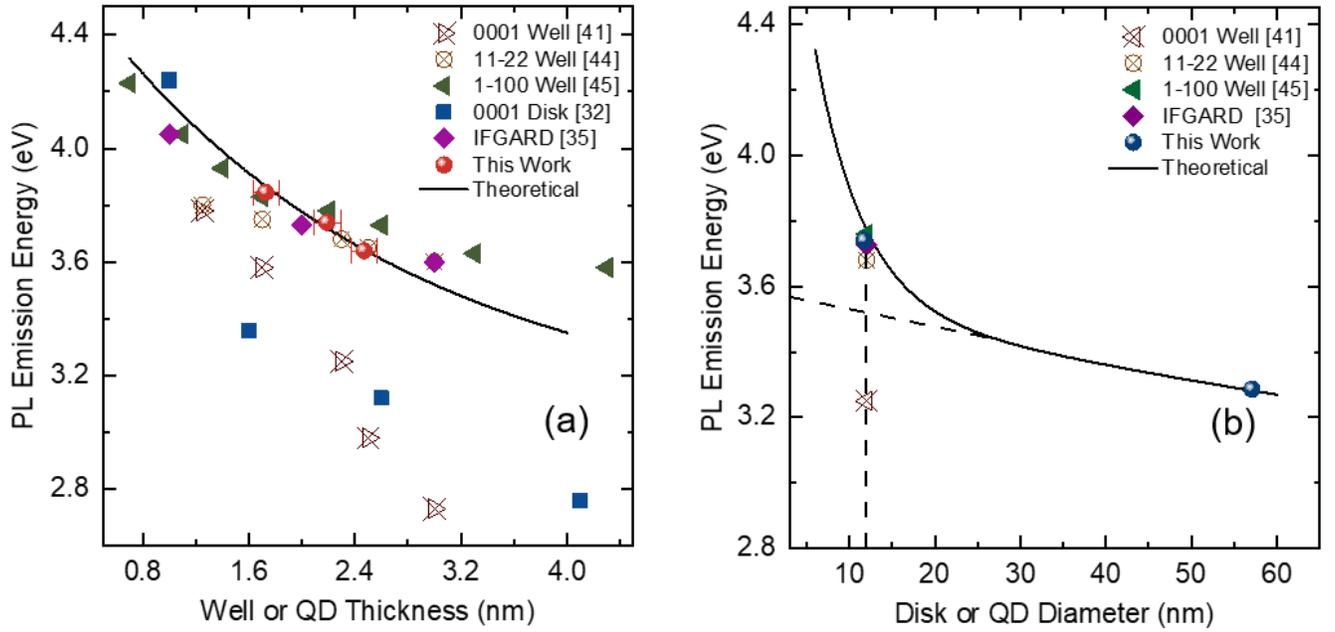

Figure 7: (a) Thickness-dependence of experimentally-obtained PL emission-energies (points) of GaN quantum wells, quantum disks and NWQDs, compared against exciton emission-energies obtained from single-band effective mass calculations (solid line). (b) The same data, and theoretically-obtained plot, versus the diameter of quantum disks and NWQDs.

As the bands bend and shift with the change of $h$ and $D$, the energy-states of electrons and holes are also altered. This manifests in the energy-shifts of the PL emission peaks. In Fig. 7 (a), the solid line plot shows the calculated blue-shift of the PL peak with reducing $h$, while $D$ (= 12 nm) is kept constant. Here, we have considered the difference between the ground-state energies of the confined electron and hole as the theoretical PL emission energy. The experimentally obtained emission peaks (Figs. 5 (a) - 5 (c)), from the NWQDs, lie precisely on the theoretically calculated plot. In the same figure, we also show experimental data points for PL emission energies, recorded (at 10 K) in earlier reports, for quantum wells grown on polar (0001) [32], semi-polar (11-22) [44] and non-polar (1-100) [45] substrates, and quantum disks grown on polar substrates, with and without IFGARD. A comparison of these data points, for any particular value of $h$, shows that emission energies of the polar structures (both wells and disks) are red-shifted w.r.t. their non-polar counterparts (or disks with IFGARD), which is due to the quantum-confined stark effect (QCSE). Interestingly, the emission energies of our NWQDs match very well with the latter. Thus, it may be inferred that despite being grown on polar substrates, there is no influence of the QCSE in the PL emitted by the NWQDs. This indicates that the built-in potential, due to spontaneous and piezoelectric polarization, is effectively suppressed. As the built-in electric field was earlier observed to grow with reducing $h$ (Fig. 6 (e)), the suppression of the polarization field is necessarily mediated by the strong reduction of $D$ (Fig. 6 (f)). Thus, approaching the strong quantum-confinement regime by shrinking the QD-diameter, as opposed to reducing the QD-height, provides the desirable advantage of suppressing the built-in potential.

It is evident that both reduction of polarization charges, and increase of quantum confinement, are responsible for the PL blue-shift of the NWQDs, with decreasing $D$. Here, we analyze the relative contribution of the two factors. For a constant $h$ (= 2 nm), Fig. 7(b) shows the calculated blue-shift of the PL peak with reducing $D$. As $D$ reduces, the blue-shift first increases linearly (down to $D \sim 25$ nm), but subsequently increases exponentially. While the linear increase is due to the suppression of the built-in potential, the exponential increase below $D \sim 25$ nm is due to the additional contribution from the strong radial-confinement. The NWQD PL peak of sample II ($h_{QD}^{(II)}$ = 2.19 nm , $D_{QD}^{(II)}$ = 11.21 nm) and the quantum-disk PL peak of sample V ($h_{QDisk}^{(V)}$ = 2.03 nm, $D_{QDisk}^{(V)}$ = 57.08 nm), lie on opposite extremes of the simulated plot. PL peak positions of the polar, semi-polar and non-polar quantum wells, as well as that for the quantum disk with IFGARD, are re-plotted in Fig. 7 (b). All these structures have approximately similar value of $h$, as that of the NWQDs of sample II. Remarkably, the peak positions corresponding to the non-polar quantum well, the quantum disk with IFGARD and sample II show nearly-exact coincidence. On the other hand, the emission energies corresponding to the polar (semi-polar) quantum wells are red-shifted by 389 meV (119 meV). By extrapolating the linear region of the simulated plot to the small-$D$ limit, we estimate the blue-shift due to suppression of polarization charges to be 236 meV, while that due to strong radial-confinement to be 218 meV. This convincingly establishes the inference that built-in potential due to the spontaneous and piezoelectric polarizations are almost completely suppressed, in the NWQDs grown by our approach.

Conclusion:

In conclusion, we reported the development of a single-step epitaxial growth technique, which allows nanowire-supported quantum dots of GaN to be scaled down to the strong quantum-confinement limit. In these ultra-scaled GaN NWQDs, the spontaneous-and-piezoelectric-polarization-induced built-in electric field, inherent to III-nitride (hetero)systems, is almost completely suppressed. SEM and HRTEM confirmed the formation of ultra-thin GaN quantum wires (QWs), supported by AlN-capped GaN nanowires, due to selective thermal decomposition of a GaN extension, deposited after AlN-capping. The QDs were then formed by depositing a tiny AlN/GaN/AlN heterostructure at the apices of the QWs. Corroborating the QD photoluminescence emission with ground-state exciton-energies obtained from a single-band effective mass calculation, and comparing them with previously-reported experimental PL data from quantum wells, quantum disks and quantum dots, we established that reduction in the QD-diameter, rather than the QD-height, allows the built-in potential of (0001)-oriented QDs to be suppressed. The capabilities achieved through this work bode extremely well for GaN NWQDs to be exploited for development of UV light-emitting-devices and single photon sources.

**ASSOCIATED CONTENT:**

The detailed calculation of the total built-in potential of QD.

**AUTHOR INFORMATION:**

**Corresponding Authors:**

*E-mail: suddho@phy.iitb.ac.in

*E-mail: laha@ee.iitb.ac.in

ACKNOWLEDGEMENTS:

All authors acknowledge the financial support from the Science and Engineering Research Board (SERB) (project no: CRG/2018/001405) of the Department of Science and Technology (DST), and the Ministry of Electronics and Information Technology (MeitY), of the Govt. of India (GoI). The authors also thank the IIT Bombay Nanofabrication Facility (IITBNF) for the technical support, towards the execution of this project. The authors thank Prof. Sandip Ghosh, faculty of the Department of Condensed Matter Physics & Materials Science, TIFR, India, for giving access to the low-temperature PL measurement facility at TIFR. Swagata Bhunia acknowledges the financial assistant of the University Grants Commission, GoI.

REFERENCES:

1. Sarwar, A. T. M. G.; May, B. J.; Deitz, J. I.; Grassman, T. J.; McComb, D. W.; Myers, R. C., Tunnel junction enhanced nanowire ultraviolet light emitting diodes. *Appl. Phys. Lett.* **2015,** *107* (10), 101103.
2. Sarwar, A. T. M. G.; May, B. J.; Chisholm, M. F.; Duscher, G. J.; Myers, R. C., Ultrathin GaN quantum disk nanowire LEDs with sub-250 nm electroluminescence. *Nanoscale* **2016,** *8* (15), 8024-8032.
3. Le, B. H.; Liu, X.; Tran, N. H.; Zhao, S.; Mi, Z., An electrically injected AlGaN nanowire defect-free photonic crystal ultraviolet laser. *Opt. Express* **2019,** *27* (4), 5843-5850.
4. Wu, Y.; Liu, X.; Wang, P.; Laleyan, D. A.; Sun, K.; Sun, Y.; Ahn, C.; Kira, M.; Kioupakis, E.; Mi, Z., Monolayer GaN excitonic deep ultraviolet light emitting diodes. *Appl. Phys. Lett.* **2020,** *116* (1), 013101.
5. Choi, K.; Kako, S.; Holmes, M. J.; Arita, M.; Arakawa, Y., Strong exciton confinement in site-controlled GaN quantum dots embedded in nanowires. *Appl. Phys. Lett.* **2013,** *103* (17), 171907.
6. Akasaki, I.; Amano, H., Crystal Growth and Conductivity Control of Group III Nitride Semiconductors and Their Application to Short Wavelength Light Emitters. *Jpn. J. Appl. Phys.* **1997,** *36* (Part 1, No. 9A), 5393-5408.
7. Shan, W.; Little, B. D.; Fischer, A. J.; Song, J. J.; Goldenberg, B.; Perry, W. G.; Bremser, M. D.; Davis, R. F., Binding energy for the intrinsic excitons in wurtzite GaN. *Physical Review B* **1996,** *54* (23), 16369-16372.
8. Mishra, U. K.; Shen, L.; Kazior, T. E.; Wu, Y. F., GaN-Based RF Power Devices and Amplifiers. *Proc. IEEE* **2008,** *96* (2), 287-305.
9. Kako, S.; Santori, C.; Hoshino, K.; Götzinger, S.; Yamamoto, Y.; Arakawa, Y., A gallium nitride single-photon source operating at 200 K. *Nature Materials* **2006,** *5* (11), 887-892.
10. Davies, S. C.; Mowbray, D. J.; Ranalli, F.; Parbrook, P. J.; Wang, Q.; Wang, T.; Yea, B. S.; Sherliker, B. J.; Halsall, M. P.; Kashtiban, R. J.; Bangert, U., Optical and microstructural studies of InGaN/GaN quantum dot ensembles. *Appl. Phys. Lett.* **2009,** *95* (11), 111903.
11. Renard, J.; Kandaswamy, P. K.; Monroy, E.; Gayral, B., Suppression of nonradiative processes in long-lived polar GaN/AlN quantum dots. *Appl. Phys. Lett.* **2009,** *95* (13), 131903.
12. Zhang, M.; Bhattacharya, P.; Guo, W., InGaN/GaN self-organized quantum dot green light emitting diodes with reduced efficiency droop. *Appl. Phys. Lett.* **2010,** *97* (1), 011103.
13. Wu, Y.-R.; Lin, Y.-Y.; Huang, H.-H.; Singh, J., Electronic and optical properties of InGaN quantum dot based light emitters for solid state lighting. *J. Appl. Phys.* **2009,** *105* (1), 013117.
14. Brown, J.; Wu, F.; Petroff, P. M.; Speck, J. S., GaN quantum dot density control by rf-plasma molecular beam epitaxy. *Appl. Phys. Lett.* **2004,** *84* (5), 690-692.
15. Tamariz, S.; Callsen, G.; Grandjean, N., Density control of GaN quantum dots on AlN single crystal. *Appl. Phys. Lett.* **2019,** *114* (8), 082101.
16. Bardoux, R.; Guillet, T.; Lefebvre, P.; Taliercio, T.; Bretagnon, T.; Rousset, S.; Gil, B.; Semond, F., Photoluminescence of single $\mathrm{GaN}/\mathrm{AlN}$ hexagonal quantum dots on $\mathrm{Si}(111)$: Spectral diffusion effects. *Physical Review B* **2006,** *74* (19), 195319.
17. Yu, J.; Hao, Z.; Li, L.; Wang, L.; Luo, Y.; Wang, J.; Sun, C.; Han, Y.; Xiong, B.; Li, H., Influence of dislocation density on internal quantum efficiency of GaN-based semiconductors. *AIP Advances* **2017,** *7* (3), 035321.
18. Kim, K. S.; Hong, C. H.; Lee, W. H.; Kim, C. S.; Cha, O. H.; Yang, G. M.; Suh, E. K.; Lim, K. Y.; Lee, H. J.; Cho, H. K.; Lee, J. Y.; Seo, J. M., Fabrication and Characterization of InGaN Nano-scale Dots for Blue and Green LED Applications. *MRS Internet J. Nitride Semicond. Res.* **2000,** *5* (1), 880-886.
19. Nakaoka, T.; Kako, S.; Arakawa, Y., Unconventional quantum-confined Stark effect in a single $\mathrm{GaN}$ quantum dot. *Physical Review B* **2006,** *73* (12), 121305.
20. Choi, K.; Arita, M.; Kako, S.; Arakawa, Y., Site-controlled growth of single GaN quantum dots in nanowires by MOCVD. *J. Cryst. Growth* **2013,** *370*, 328-331.
21. Renard, J.; Songmuang, R.; Bougerol, C.; Daudin, B.; Gayral, B., Exciton and Biexciton Luminescence from Single GaN/AlN Quantum Dots in Nanowires. *Nano Lett.* **2008,** *8* (7), 2092-2096.
22. Hersee, S. D.; Rishinaramangalam, A. K.; Fairchild, M. N.; Zhang, L.; Varangis, P., Threading defect elimination in GaN nanowires. *J. Mater. Res.* **2011,** *26* (17), 2293-2298.
23. Calleja, E.; Sánchez-García, M. A.; Sánchez, F. J.; Calle, F.; Naranjo, F. B.; Muñoz, E.; Jahn, U.; Ploog, K., Luminescence properties and defects in GaN nanocolumns grown by molecular beam epitaxy. *Physical Review B* **2000,** *62* (24), 16826-16834.
24. Zadeh, I. E.; Elshaari, A. W.; Jöns, K. D.; Fognini, A.; Dalacu, D.; Poole, P. J.; Reimer, M. E.; Zwiller, V., Deterministic Integration of Single Photon Sources in Silicon Based Photonic Circuits. *Nano Lett.* **2016,** *16* (4), 2289-2294.
25. Himwas, C.; den Hertog, M.; Bellet-Amalric, E.; Songmuang, R.; Donatini, F.; Si Dang, L.; Monroy, E., Enhanced room-temperature mid-ultraviolet emission from AlGaN/AlN Stranski-Krastanov quantum dots. *J. Appl. Phys.* **2014,** *116* (2), 023502.

26. Himwas, C.; Songmuang, R.; Le Si, D.; Bleuse, J.; Rapenne, L.; Sarigiannidou, E.; Monroy, E., Thermal stability of the deep ultraviolet emission from AlGaN/AlN Stranski-Krastanov quantum dots. *Appl. Phys. Lett.* **2012,** *101* (24), 241914.
27. Holmes, M. J.; Kako, S.; Choi, K.; Arita, M.; Arakawa, Y., Single Photons from a Hot Solid-State Emitter at 350 K. *ACS Photonics* **2016,** *3* (4), 543-546.
28. Djavid, M.; Mi, Z., Enhancing the light extraction efficiency of AlGaN deep ultraviolet light emitting diodes by using nanowire structures. *Appl. Phys. Lett.* **2016,** *108* (5), 051102.
29. Beeler, M.; Lim, C. B.; Hille, P.; Bleuse, J.; Schörmann, J.; de la Mata, M.; Arbiol, J.; Eickhoff, M.; Monroy, E., Long-lived excitons in GaN/AlN nanowire heterostructures. *Physical Review B* **2015,** *91* (20), 205440.
30. Frost, T.; Banerjee, A.; Sun, K.; Chuang, S. L.; Bhattacharya, P., InGaN/GaN Quantum Dot Red $(\lambda=630~{\rm nm})$ Laser. *IEEE J. Quantum Electron.* **2013,** *49* (11), 923-931.
31. Holmes, M. J.; Choi, K.; Kako, S.; Arita, M.; Arakawa, Y., Room-Temperature Triggered Single Photon Emission from a III-Nitride Site-Controlled Nanowire Quantum Dot. *Nano Lett.* **2014,** *14* (2), 982-986.
32. Renard, J.; Songmuang, R.; Tourbot, G.; Bougerol, C.; Daudin, B.; Gayral, B., Evidence for quantum-confined Stark effect in GaN/AlN quantum dots in nanowires. *Physical Review B* **2009,** *80* (12), 121305.
33. Pelekanos, N. T.; Dialynas, G. E.; Simon, J.; Mariette, H.; Daudin, B., GaN quantum dots: from basic understanding to unique applications. *Journal of Physics: Conference Series* **2005,** *10*, 61-68.
34. Hille, P.; Müßener, J.; Becker, P.; de la Mata, M.; Rosemann, N.; Magén, C.; Arbiol, J.; Teubert, J.; Chatterjee, S.; Schörmann, J.; Eickhoff, M., Screening of the quantum-confined Stark effect in AlN/GaN nanowire superlattices by germanium doping. *Appl. Phys. Lett.* **2014,** *104* (10), 102104.
35. Schlichting, S.; Hönig, G. M. O.; Müßener, J.; Hille, P.; Grieb, T.; Westerkamp, S.; Teubert, J.; Schörmann, J.; Wagner, M. R.; Rosenauer, A.; Eickhoff, M.; Hoffmann, A.; Callsen, G., Suppression of the quantum-confined Stark effect in polar nitride heterostructures. *Communications Physics* **2018,** *1* (1), 48.
36. Zettler, J. K.; Corfdir, P.; Hauswald, C.; Luna, E.; Jahn, U.; Flissikowski, T.; Schmidt, E.; Ronning, C.; Trampert, A.; Geelhaar, L.; Grahn, H. T.; Brandt, O.; Fernández-Garrido, S., Observation of Dielectrically Confined Excitons in Ultrathin GaN Nanowires up to Room Temperature. *Nano Lett.* **2016,** *16* (2), 973-980.
37. Brockway, L.; Pendyala, C.; Jasinski, J.; Sunkara, M. K.; Vaddiraju, S., A Postsynthesis Decomposition Strategy for Group III–Nitride Quantum Wires. *Crystal Growth & Design* **2011,** *11* (10), 4559-4564.
38. Bhunia, S.; Sarkar, R.; Nag, D.; Ghosh, K.; Khiangte, K. R.; Mahapatra, S.; Laha, A., Decomposition Resilience of GaN Nanowires, Crested and Surficially Passivated by AlN. *Crystal Growth & Design* **2020,** *20* (8), 4867-4874.
39. Auzelle, T. GaN/AlN nanowires : nucleation, polarity and quantum heterostructures
Nanofils de GaN/AlN : nucléation, polarité et hétérostructures quantiques. Université Grenoble Alpes, 2015.
40. Schneider, C. A.; Rasband, W. S.; Eliceiri, K. W., NIH Image to ImageJ: 25 years of image analysis. *Nat Methods* **2012,** *9* (7), 671-675.
41. Adelmann, C.; Sarigiannidou, E.; Jalabert, D.; Hori, Y.; Rouvière, J. L.; Daudin, B.; Fanget, S.; Bru-Chevallier, C.; Shibata, T.; Tanaka, M., Growth and optical properties of GaN/AlN quantum wells. *Appl. Phys. Lett.* **2003,** *82* (23), 4154-4156.
42. Andreev, A. D.; O'Reilly, E. P., Theory of the electronic structure of GaN/AlN hexagonal quantum dots. *Physical Review B* **2000,** *62* (23), 15851-15870.
43. Andreev, A. D.; Downes, J. R.; Faux, D. A.; O'Reilly, E. P., Strain distributions in quantum dots of arbitrary shape. *J. Appl. Phys.* **1999,** *86* (1), 297-305.
44. Lahourcade, L.; Kandaswamy, P. K.; Renard, J.; Ruterana, P.; Machhadani, H.; Tchernycheva, M.; Julien, F. H.; Gayral, B.; Monroy, E., Interband and intersubband optical characterization of semipolar (112̄2)-oriented GaN/AlN multiple-quantum-well structures. *Appl. Phys. Lett.* **2008,** *93* (11), 111906.
45. Grenier, V.; Finot, S.; Jacopin, G.; Bougerol, C.; Robin, E.; Mollard, N.; Gayral, B.; Monroy, E.; Eymery, J.; Durand, C., UV Emission from GaN Wires with m-Plane Core–Shell GaN/AlGaN Multiple Quantum Wells. *ACS Applied Materials & Interfaces* **2020,** *12* (39), 44007-44016.